\documentclass[aps,pra,showpacs,eqsecnum,twocolumn]{revtex4}
\usepackage{graphicx}
\usepackage{bm}
\usepackage{color}

\begin{document}
\preprint{}
\title{Nonclassicality of coherent states: Entanglement of joint statistics}
\author{Alfredo Luis}
\email{alluis@fis.ucm.es}
\homepage{http://www.ucm.es/info/gioq}
\author{Laura Monroy} 
\affiliation{Departamento de \'{O}ptica, Facultad de Ciencias
F\'{\i}sicas, Universidad Complutense, 28040 Madrid, Spain}
\date{\today}

\begin{abstract}
Simple joint measurements of pairs of observables reveal that states considered universally as 
classical-like, such as SU(2) spin coherent states, Glauber coherent states, and thermal states 
are actually nonclassical. We show that this holds because we can find a joint measurement the 
statistics of which is not separable. Eventually this may be extended to all states different from the 
maximally mixed state.
\end{abstract}


\maketitle

\section{Introduction}

Quantumness is the {\em raison d'\^ etre} of the quantum theory as well as the resource behind the 
quantum-technology revolution, as exemplified via entanglement and Bell's tests. In the conventional 
approach, quantumness holds for a limited class of states, difficult to generate and preserve in practice. 

Nonclassical effects always emerge as  the impossibility of confining randomness of two or more 
variables within probability distributions  \cite{thisone}.  For example, this is actually the case of the 
celebrated quantum tests of the Bell type \cite{WMB,AF,TMN}.  This includes as a particular case 
mainstream tests  such as the failure of the Glauber-Sudarshan $P$ function to be a true probability 
density \cite{P}. 

In this work we provide a rather new perspective by showing that states considered universally as 
classical-like, such as SU(2) spin coherent states, Glauber coherent states, and thermal states, are 
actually nonclassical. Eventually this can be extended to all states except the maximally mixed state. 
We show that this holds via the new idea of entanglement of joint statistics.  

To show this we consider the simultaneous measurement of two compatible observables in an 
enlarged system-apparatus space,  that provides complete information about the statistics of two 
incompatible system observables. This is to say that we can recover their exact individual statistics after a suitable 
data inversion applied to the corresponding observed marginal distributions \cite{WMM}. Then we 
apply the data inversion to the joint statistics. In classical physics this always leads to the joint 
statistics of the corresponding system observables, a {\it bona fide} probability distribution. We 
show that this holds because in classical physics all joint distributions are separable, so the 
inversion of the joint distribution works equally well as the inversion of the marginals. 

However, in quantum physics this is not longer the case, and the inversion can lead to pathological joint 
distributions that are not probabilities. In such a case we say that  the state is nonclassical. We 
describe the general procedure in Sec. II. We apply it to the qubit case in Sec. III and to Glauber 
quadrature coherent states and thermal states in Sec. IV via quadrature measurements in double 
homodyne detection.

\section{Basic settings}

Nonclassicality cannot be a single-observable property since within classical physics it is always 
possible to reproduce exactly the statistics of any quantum observable. Nonclassical effects can 
only emerge when addressing the joint statistics of multiple observables,  especially if they are 
incompatible. Let us show how from two different perspectives: probability distributions and 
characteristic functions. 

\subsection{Probability distributions}

In the most general case, joint measurements require the coupling of the system space ${\cal H}_s$ 
with auxiliary degrees of freedom ${\cal H}_a$. We consider the simultaneous measurement of two 
compatible observables, $\tilde{X}$ and $\tilde{Y}$, in the enlarged space ${\cal H}_s \otimes {\cal H}_a$  
with outcomes $x$ and $y$, respectively, and joint probability $\tilde{p}_{X,Y} (x,y)$. Since this corresponds 
to the statistics or a real measurement we have that  $\tilde{p}_{X,Y} (x,y)$ is a well-behaved probability 
distribution. The corresponding marginal distributions are 
\begin{equation}
\label{itn}
\tilde{p}_X (x)  = \sum_y \,\tilde{p}_{X,Y} (x,y), \quad \tilde{p}_Y (y) = \sum_x \,\tilde{p}_{X,Y} (x,y) ,
\end{equation}
where we are assuming a discrete range for  $x$ and $y$ without loss  of generality. We assume that 
these marginals provide complete information about two system observables in the system space 
${\cal H}_s$, say $X$ and $Y$, respectively, maybe incompatible. This is to say that their probability 
distributions $p_A  (a)$ for $A=X,Y$ and $a, a^\prime=x,y$ can be retrieved from the observed 
marginals  $\tilde{p}_A  (a)$ as
\begin{equation}
\label{inv}
p_A (a) = \sum_{a^\prime} \mu_A (a, a^\prime )\; \tilde{p}_A  (a^\prime) ,
\end{equation}
where the functions $\mu_A (a, a^\prime )$ are completely known as far as we know the measurement being 
performed and the initial state of the auxiliary degrees of freedom ${\cal H}_a$. 
 We stress that relation (\ref{inv}) is an assumption that holds or not depending on the observable $A$, the 
measurement performed, and the initial state of the ancilla. Whenever the inversion is possible, the functions
$\mu_A (a, a^\prime )$ can be easily determined by imposing (\ref{inv}) for arbitrary states of the system being 
observed. 

The key idea is to extend  this inversion (\ref{inv}) from the marginals to the complete joint distribution to obtain a 
joint distribution $p_{X,Y} (x,y)$ for the $X,Y$ variables in the system state \cite{WMM}:
\begin{equation}
\label{nti}
p_{X,Y} (x,y)  = \sum_{x^\prime, y^\prime}\, \mu_X (x, x^\prime)\, \mu_{Y} (y, y^\prime )  
\,\tilde{p}_{X,Y} (x^\prime , y^\prime ) .
\end{equation}
This is actually a definition of $p_{X,Y} (x,y)$ motivated by the classical case, where after Eq. (\ref{inv}) the  kernels  
$\mu_A (a, a^\prime )$ are actually independent conditional probabilities of getting $a$ given $a^\prime$, with 
\begin{equation}
\sum_{a} \mu_A (a, a^\prime ) = 1.
\end{equation}
In particular this implies that the kernels in Eq. (\ref{nti}) must be the same introduced in Eqs. (\ref{inv}) since 
$p_{X,Y} (x,y)$ must give the correct marginals 
\begin{equation}
p_X (x) = \sum_y p_{X,Y} (x,y)  , \quad p_Y (y) = \sum_x p_{X,Y} (x,y)  .
\end{equation}
Parallels can be drawn with the construction joint probability distributions via the inversion of moments \cite{UD}. 

\subsection{Characteristic functions}

An alternative approach can be formulated  in terms of characteristic functions defined as usual 
as the Fourier transform of the probability distributions, assuming now a continuous range for $a$, 
\begin{equation}
C_A ( u  ) = \int da\, e^{i a u} \;p_A (a) = \langle e^{iuA} \rangle , 
\end{equation}
which can be inverted in the form 
\begin{equation}
p_A ( a ) = \frac{1}{2\pi} \int du\, e^{-i a u}\, C_A (u) .
\end{equation}
Since both $C_A (u)$ and $p_A (a)$ contain full information about the statistics of $A$ the characteristic 
function can equally well serve for our purposes.

The simultaneous measurement of $\tilde{X}$, $\tilde{Y}$ leads to a joint characteristic function 
\begin{equation}
\tilde{C}_{X,Y} (u,v) = \langle e^{i\left ( u \tilde{X} + v \tilde{Y} \right )} \rangle = \int dx dy e^{i\left ( u x + v y \right )} 
\tilde{p}_{X,Y} (x,y) ,
\end{equation}
from which two marginal characteristics can be derived for each observable rather simply as
\begin{equation}
\tilde{C}_{X} (u) = \tilde{C}_{X,Y} (u,0), \qquad \tilde{C}_{Y} (v) = \tilde{C}_{X,Y} (0,v).
\end{equation}
In many interesting practical settings, such as the one to be examined in Sec. IV, the observed $\tilde{C}_A ( u )$ 
and true $C_A ( u )$ characteristics  are simply related in the form 
\begin{equation}
\label{CHC}
\tilde{C}_A ( u ) = H_A (u) C_A ( u  ) ,
\end{equation}
where $H_A (u)$ is an instrumental function, which is assumed to be known as far as we know the details of the 
measurement being performed. This is the case of linear shift invariant systems where $H$ is the frequency response 
of the system, or the optical transfer function in classical imaging optics. That is to say that $\tilde{p}_A (a)$ is the result 
of convolving $p_A (a)$ with  the impulse response function, which is the Fourier transform of $H$. Note that 
$H_A(0)=1$ by normalization of probability distributions. 

Assuming that $ H_A (u)$ has no zeros, as it will be our case here, the analog of the inversion (\ref{inv}) is after
Eq. (\ref{CHC}) simply 
\begin{equation}
C_A ( u  ) = \tilde{C}_A ( u ) / H_A (u). 
\end{equation}
Applying the inversion to the joint statistics we get 
\begin{equation}
\label{fc}
C_{X,Y} ( u,v ) = \frac{\tilde{C}_{X,Y} (u,v) }{H_X (u) H_Y (v)} ,
\end{equation}
as a particular counterpart of Eq. (\ref{nti}). The question is whether the so inferred characteristic $C_{X,Y} ( u,v ) $ leads 
to a true probability distribution $p_{X,Y} (x,y) $ via Fourier inversion:
\begin{equation}
\label{fi}
p_{X,Y} (x,y) = \frac{1}{(2 \pi)^2} \int du\, dv\, e^{-i\left ( u x + v y \right )}\; C_{X,Y} (u,v) ,
\end{equation}
i. e., whether the integral exists and  $p_{X,Y} (x,y) $  is nonnegative, as it is always the case in classical physics as 
shown next.

\subsection{Classical physics}

Let us show that in classical physics these inversion procedures (\ref{nti}) and (\ref{fc}) always lead to a \textit{bona fide} 
probability distribution $p_{X,Y} (x,y)$. Classically the state of the system can be completely described by a legitimate 
probability distribution  $p_j$, where index $j$ runs over all admissible states $\bm{\lambda}_j$ for the system. This is 
the corresponding phase space, assumed to form a discrete set for simplicity and without loss of generality.  There is no 
limit to the number of points  $\bm{\lambda}_j$ so it may approach a continuum if necessary. 

So the observed joint statistics can be always expressed as
\begin{equation}
\label{jj}
\tilde{p}_{X,Y} (x,y) = \sum_j\, p_j\, \tilde{X} (x | \bm{\lambda}_j)\;  \tilde{Y} (y | \bm{\lambda}_j) ,
\end{equation}
where  $\tilde{A}(a | \bm{\lambda}_j) $ is the conditional probability that the observable $\tilde{A}$ 
takes the value $a$ when the system state is $\bm{\lambda}_j$. By definition, phase-space points 
$\bm{\lambda}_j$ have definite values for every observable so the factorization $\tilde{X} (x | \bm{\lambda}_j)  
\tilde{Y} (y | \bm{\lambda}_j)$ holds. Strictly speaking they are the product of delta functions. 
Applying Eq. (\ref{inv}) we get the conditional probabilities for the system variables  
\begin{equation}
\label{rel}
 A( a | \lambda_j ) = \sum_{a^\prime} \,\mu_A (a,a^\prime )\;  \tilde{A} ( a^\prime | \bm{\lambda}_j ) .
\end{equation}
Thus, because of the separable form (\ref{jj}) we readily get from Eqs. (\ref{nti}) and (\ref{rel}) that the result of 
the inversion is  the actual joint distribution for $X$ and $Y$
\begin{equation}
\label{legi}
p_{X,Y} (x,y)  = \sum_j  \,p_j\, X(x | \bm{\lambda}_j)\;  Y (y | \bm{\lambda}_j ) ,
\end{equation}
and therefore a legitimate statistics. Thus, lack of positivity or any other pathology of the retrieved joint 
distribution $p_{X,Y} (x,y) $ is then a signature of nonclassical behavior. 

Similarly, the procedure outlined above in terms of characteristic functions leads always in classical physics to 
a {\it bona fide} distribution. This is because  the observed characteristics is always separable as the Fourier 
transform of Eq. (\ref{jj}) 
\begin{equation}
\tilde{C}_{X,Y} (u,v)  = \sum_j  \,p_j\, \tilde{C}_X (u | \bm{\lambda}_j)\,  \tilde{C}_Y  (v | \bm{\lambda}_j ) ,
\end{equation}
where $\tilde{C}_A (u | \bm{\lambda}_j)$ are the corresponding conditional characteristics. Then, after     
Eq. (\ref{fc}) we get also a separable joint characteristics for system variables $X$, $Y$
\begin{equation}
C_{X,Y} (u,v)  = \sum_j  \,p_j \,C_X (u | \bm{\lambda}_j)\, C_Y  (v | \bm{\lambda}_j ) ,
\end{equation}
that leads via Fourier transform to the same legitimate distribution in Eq. (\ref{legi}). 

\section{Qubit example}

Let us focus on the qubit as the simplest quantum system ${\cal H}_s$. The most general state of the qubit is 
\begin{equation}
\label{ss}
\rho = \frac{1}{2} \left ( \sigma_0 + \bm{s} \cdot \bm{\sigma}  \right ) , \quad |\bm{s} | \leq 1, 
\end{equation}
where $\bm{s}$ is a three-dimensional real vector with $| \bm{s}| \leq 1$, $\sigma_0$ is the $2 \times 2$ identity
matrix, and $\bm{\sigma}$ are the Pauli matrices. The task is finding for every $\rho$ a suitable measurement so 
that the inversion (\ref{nti})  of the observed statistics  $\tilde{p}_{X,Y} (x,y)$ cannot be a probability distribution. 
To this end, we will use that any measurement performed in the enlarged space ${\cal H}_s  \otimes {\cal H}_a$ 
can be conveniently described by a positive operator-valued measure in ${\cal H}_s$ 
\begin{equation}
\label{tpovm}
\tilde{\Delta}_{X,Y} (x,y) = \frac{1}{4} \left ( \sigma_0 + \bm{\eta} (x,y) \cdot \bm{\sigma}  \right )  .
\end{equation}
Positivity and normalization require that 
\begin{equation}
\tilde{\Delta}_{X,Y} (x,y) \geq 0, \qquad \sum_{x,y} \tilde{\Delta}_{X,Y} (x,y) = \sigma_0 ,
\end{equation}
so that 
\begin{equation}
| \bm{\eta} (x,y) |  \leq 1 , \qquad \sum_{x,y}  \bm{\eta} (x,y) = \bm{0} .
\end{equation}
The corresponding statistics is 
\begin{equation}
\label{ooee}
\tilde{p}_{X,Y} (x,y)= \mathrm{tr} \left [ \rho \tilde{\Delta}_{X,Y} (x,y) \right ] = \frac{1}{4} \left ( 1 +  \bm{\eta} (x,y) \cdot 
\bm{s}  \right ) ,
\end{equation}
and naturally 
\begin{equation}
\tilde{p}_{X,Y} (x,y) \geq 0, \qquad  \sum_{x,y} \tilde{p}_{X,Y} (x,y) = 1.
\end{equation}

For definiteness, let us consider the case  
\begin{equation}
\label{ge}
\bm{\eta} (x,y) = \frac{\eta}{\sqrt{3}} \left (x, y,xy \right ) ,
\end{equation}
where  $x,y = \pm 1$ and  $\eta$ is a real parameter we will assume positive without loss of generality $1 \geq \eta >0$.
Actually, for $\eta = 1$ we have that $\tilde{p}_{X,Y} (x,y)$ is a discrete and complete sampling of the SU(2) Husimi function 
for two-dimensional systems \cite{dQ}. The observed marginals are
\begin{equation}
\label{oe}
\tilde{p}_X (x)  = \frac{1}{2} \left ( 1 +  x \frac{\eta}{\sqrt{3}} s_x \right ),
\quad
\tilde{p}_Y (y)  = \frac{1}{2} \left ( 1 +  y \frac{\eta}{\sqrt{3}} s_y \right ),
\end{equation}
that provide complete information about the system observables $X = \sigma_x$, $Y= \sigma_y$ with exact statistics 
\begin{equation}
p_X (x)  = \frac{1}{2} \left ( 1 +  x s_x \right ),
\quad
p_Y (y)  = \frac{1}{2} \left ( 1 +  y s_y  \right ) .
\end{equation}
The inversion of the marginals is carried out by the functions
\begin{equation}
\mu_A \left ( a ,  a^\prime \right ) = \frac{1}{2} \left ( 1 + \frac{\sqrt{3}}{\eta}  a a^\prime \right )  ,
\end{equation}
so that the inversion of the joint distribution in Eq.  (\ref{nti}) leads to
\begin{equation}
p_{X,Y} (x,y) = \frac{1}{4} \left ( 1 + x s_x  +   y s_y + x y s_z \frac{\sqrt{3}}{\eta} \right ) .
\end{equation}

\subsection{All states different from the maximally mixed state are nonclassical}

Throughout we are free to chose the axes and the observables measured. In this spirit, using SU(2) symmetry, 
and without loss of generality, we can choose axes so that $s_x = s_y =0, s_z = |\bm{s} |$, so that 
\begin{equation}
\label{es}
p_{X,Y} (x,y) = \frac{1}{4} \left ( 1 +   x y \frac{\sqrt{3}} {\eta}| \bm{s} |  \right ) .
\end{equation}
This can take negative values for $x = - y = \pm 1$ 
\begin{equation}
\label{neg}
p_{X,Y} (\pm 1,\mp 1) = \frac{1}{4} \left ( 1 -  \frac{\sqrt{3}} {\eta} | \bm{s} |  \right )  < 0,
\end{equation}
provided that $\eta < \sqrt{3}  | \bm{s} |$. Clearly for all $\bm{s} \neq \bm{0}$ we can always chose $\eta$ 
satisfying this relation. So every state different from the identity is non classical. The identity being the 
maximally mixed state $\bm{s}=\bm{0}$.

In this regard it is worth noting that all pure states of the qubit are SU(2) coherent states \cite{cs1}.  
Because of their definition and properties they are often regarded as  the closets analogs of the Glauber coherent 
states that can exist in finite-dimensional spaces. Accordingly, since Glauber coherent states are universally 
regarded as classical, the SU(2) coherent states are reported as the most classical allowed in finite-dimensional 
systems. This is because their joint angular-momentum statistics can be described by a {\em bona fide}  classical-like 
distribution on the corresponding phase space, which is the sphere. This is discussed in great  detail in Ref. \cite{cs2} 
for example, regarding their Glauber-Sudarshan SU(2) $P$-function. This is to say that their classical-like 
resemblance refers to their angular-momentum statistical properties, although they would be nonclassical 
by their finite-dimensional nature. However we have just shown that even if we just focus on the angular-momentum 
statistics they are actually as nonclassical as any other spin state when we look beyond the $P$-function.  

\subsection{Entanglement of statistics} 

Let us provide an explicit demonstration that if $p_{X,Y}  <0$ the observed statistics (\ref{es}) cannot 
be expressed in a separable form. Separable means that there is a {\it bona fide} probability distribution 
$p_j$ so that  
\begin{equation}
\tilde{p}_{X,Y} (x,y) = \sum_j  \frac{p_j}{4} \left ( 1 +  x \frac{\eta}{\sqrt{3}} \lambda_{j,x} \right )
\left ( 1 +  y \frac{\eta}{\sqrt{3}} \lambda_{j,y} \right ),
\end{equation}
leading to 
\begin{equation}
\label{sep}
p_{X,Y} (x,y) = \sum_j  \frac{p_j}{4} \left ( 1 +  x \lambda_{j,x} \right )
\left ( 1 +  y  \lambda_{j,y} \right ),
\end{equation}
where since the phase space is an sphere  $\bm{\lambda}_j$ are three-dimensional real vector with unit modulus,
$|\bm{\lambda}_j| \leq 1$, being $\lambda_{j,x}$ and $\lambda_{j,y}$ the corresponding components. We recall 
that there is no limit to the number of vectors $\bm{\lambda}_j$. Then, if the separable form (\ref{sep}) holds we have 
after Eq. (\ref{es}) that
\begin{equation}
\label{lc}
\sum_j p_j \lambda_{j,x} \lambda_{j,y} =   \frac{\sqrt{3}} {\eta}| \bm{s} |  . 
\end{equation}
We can readily show that separability (\ref{sep}) and negativity  (\ref{neg}) are contradictory. This  is because 
$|\bm{\lambda}_j| \leq 1$ so that $\sum_j p_j \lambda_{j,x} \lambda_{j,y} \leq 1$. Thus separability implies 
$\sqrt{3}  | \bm{s} |/\eta \leq 1$ while negativity implies just the opposite $\sqrt{3}  | \bm{s} |/\eta > 1$. Therefore, negativity 
of the inferred distribution $p_{X,Y} (x,y) $ is equivalent to entanglement of the observed statistics $\tilde{p}_{X,Y} (x,y)$. 

\subsection{Practical implementation} 

Comparing  Eqs. (\ref{ss})  and  (\ref{tpovm}) with (\ref{ge}) it can be readily seen that for $\eta=1$ the elements of the 
POVM (\ref{tpovm}) are proportional to projectors on pure states with 
\begin{equation}
\label{xyz}
\langle \sigma_x \rangle = x/\sqrt{3}, \quad \langle \sigma_y \rangle = y/\sqrt{3}, \quad \langle \sigma_z \rangle = xy/\sqrt{3} .
\end{equation} 
If we write the most general pure state in the basis of eigenvectors of $\sigma_z$ as 
\begin{equation}
\label{pst}
| \psi \rangle = \pmatrix{\cos \frac{\theta}{2} \cr \sin \frac{\theta}{2} e^{i\phi} } ,
\end{equation}
we get
\begin{equation}
\label{ssp}
\langle \sigma_x \rangle = \sin \theta \cos \phi, \quad \langle \sigma_y \rangle = \sin \theta \sin \phi , \quad \langle \sigma_z \rangle = \cos \theta ,
\end{equation} 
so that the states satisfying the conditions (\ref{xyz}) can be easily found  by suitably combining $\theta$ and $\phi$ values with
\begin{equation}
\theta = \pm \theta_0  \; \mathrm{mod} \; \pi , \qquad \phi = \pm \phi_0  \; \mathrm{mod} \; \pi, 
\end{equation}
being
\begin{equation}
\tan \theta_0 = \sqrt{2}, \qquad \phi_0 = \pi/4 .
\end{equation}

The projection on these states can be easily implemented in practice in a one-photon realization of the qubit via the version of the eight-port homodyne detector schematized in Fig. 1 \cite{WA87,NFM,LP}. Let the qubit be spanned by the one-photon states $| 1,0 \rangle$ and $| 0, 1 \rangle$, where $|n_1 , n_2 \rangle$ denote photon-number states with $n_{1,2} $ photons in two field modes $a_{1,2}$. We consider these states as the eigenstates of $\sigma_z$ with eigenvalues 1 and -1, respectively. The modes $a_{1,2}$ are mixed with two further modes in vacuum as schematized in Fig. 1. The two input  beam splitters are identical, unbalanced, with real transmission and reflection coefficients $t$, $r$, 
\begin{equation}
t = \sin \frac{\theta_0}{2}, \qquad r = \cos \frac{\theta_0}{2} ,
\end{equation} 
with a relative $\pi$ phase change in the lower-side reflections. There are also two phase plates introducing phase shifts $\phi_{1,2}$ with 
\begin{equation}
  \phi_1 = - \phi_2 =  \pi/4 .
\end{equation}   
The output beam splitters are balanced,  also with real transmission and reflection coefficients and a $\pi$ phase change in the lower-side reflections. Detectors placed at the four output beams detect the exit port of the photon, so there are only four possible outcomes. The input-output relations for the complex amplitudes are, omitting for simplicity the vacuum modes that will not contribute to the final result, 
\begin{eqnarray}
& a_3 = \frac{1}{\sqrt{2}} \left ( - r a_1 + t e^{i \phi_2} a_2 \right ) , \; a_4 = \frac{1}{\sqrt{2}} \left ( - r a_1-  t e^{i \phi_2} a_2 \right ), & \nonumber \\ 
& & \\
& a_5 = \frac{1}{\sqrt{2}} \left ( t  e^{i \phi_1}  a_1 - r a_2 \right ) , \;  a_6 = \frac{1}{\sqrt{2}} \left ( t  e^{i \phi_1}  a_1 +  r a_2 \right ) ,& \nonumber \
\end{eqnarray}
where $a_j$ is the amplitude of the field mode impinging on detector $D_j$. Following the analyses in Ref. \cite{LP}  for a one-photon case,  the probability that the detector $D_j$ clicks is $p(j) = |\langle j | \psi \rangle |^2$  where the unnormalized vectors $|j \rangle$ are, following the same criterion as in Eq. (\ref{pst}) 
\begin{eqnarray}
& | 3 \rangle = \frac{1}{\sqrt{2}} \pmatrix{r \cr - t e^{-i\phi_2} }, \;  | 4 \rangle = \frac{1}{\sqrt{2}} \pmatrix{r \cr t e^{-i\phi_2} }, & \nonumber \\
 & & \\
& | 5 \rangle = \frac{1}{\sqrt{2}} \pmatrix{t \cr - r e^{-i\phi_1} }, \;  | 6 \rangle = \frac{1}{\sqrt{2}} \pmatrix{t \cr r e^{-i\phi_1} }. & \nonumber \nonumber \
\end{eqnarray}
Therefore, using all preceding equations in this section, it can be easily seen that the detectors click with the probabilities in Eqs. (\ref{ooee}) and (\ref{ge})  for $\eta=1$. More specifically $D_3$ clicks with probability $\tilde{p}_{X,Y} (-1,-1)$, detector $D_4$ clicks with probability $\tilde{p}_{X,Y} (1,1)$,  $D_5$ clicks with probability $\tilde{p}_{X,Y} (-1,1)$, and  $D_6$ clicks with probability $\tilde{p}_{X,Y} (1,-1)$. 

\begin{figure}
\centering
\includegraphics[width=6cm]{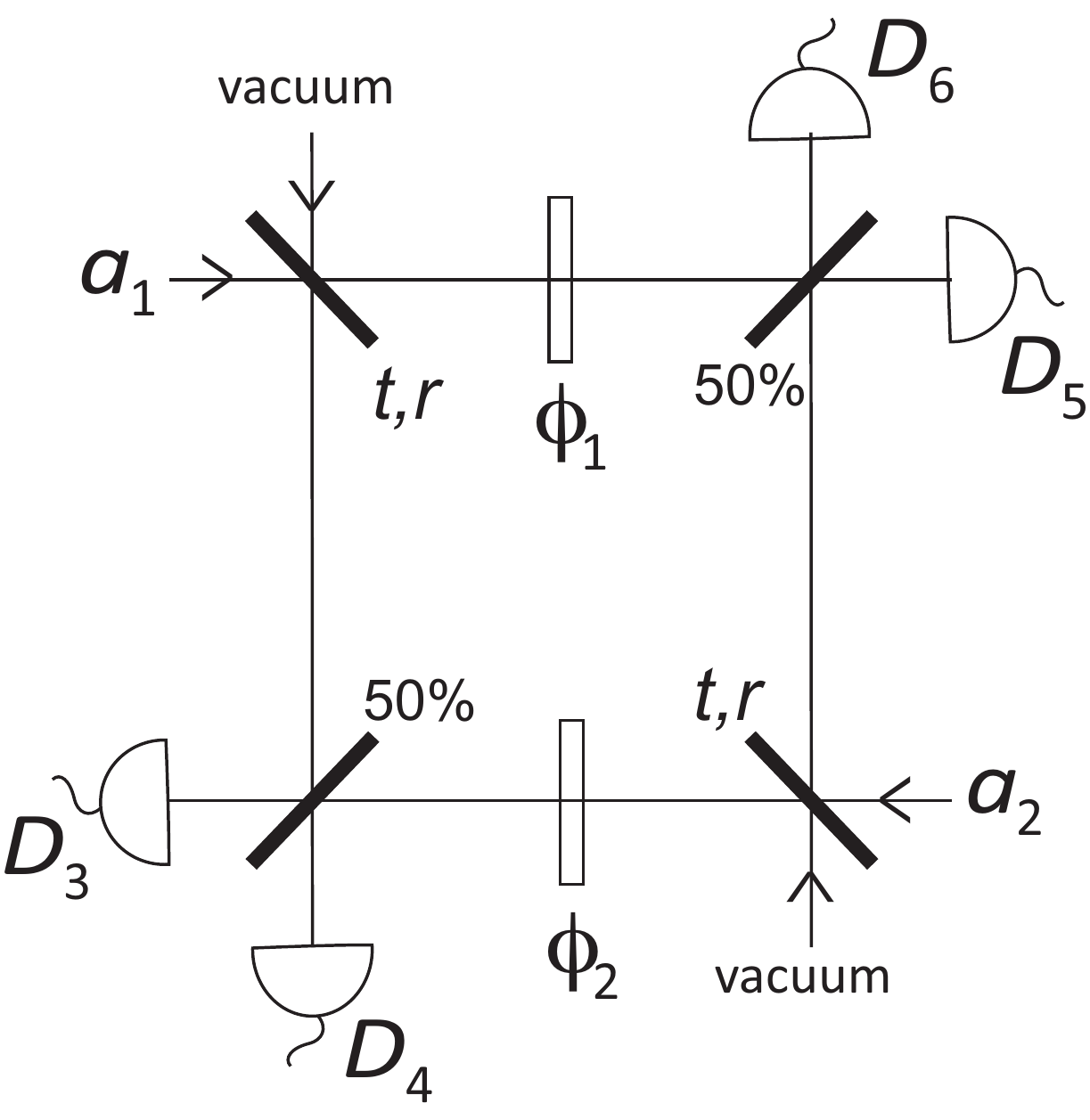}
\label{experimento}
\caption{Eight-port homodyne detector.}
\end{figure}

\subsection{Extension to larger dimension}

This analysis  may  be extended to systems in Hilbert spaces of arbitrary dimension. For pure states $| \psi \rangle$ 
this can be readily done by focusing on the two-dimensional subspace spanned by the pair $| \psi \rangle$, 
$| \psi_\perp \rangle$, where $| \psi_\perp \rangle$ is any state orthogonal to $| \psi \rangle$. We may then
define $\sigma_z = |\psi \rangle \langle  \psi |- | \psi_\perp \rangle \langle  \psi_\perp |$ and accordingly for the other 
Pauli matrices. For mixed states we may focus on their projection on any two-dimensional 
space that can be regarded as the marginal distribution of a larger statistics. Alternatively, we may deal with dichotomic 
observables, such as parity or any other on/off detectors \cite{par,SVA}. 

\section{Unbalanced double homodyne detection}

Next we will demonstrate that there is a simple practical procedure leading to pathological quadrature joint statistics 
for Glauber coherent states and thermal states. 

\subsection{Procedure}

The experiment consists of a double homodyne detector (see Fig. 2), where the observed state  $|\psi\rangle$ is mixed with vacuum in an unbalanced beam splitter, with transmission and reflection coefficients $t$ and $r$, respectively. At the output of the beam splitter, two homodyne detectors perform the simultaneous measurement of the commuting rotated quadratures $X_{1,\theta}$ and $Y_{2,\theta}$ in the corresponding modes, where $\theta$ is the phase of the local oscillator. We understand this as a noisy simultaneous measurement of the noncommuting quadratures  $X$ and $Y$  in the  signal mode in state  $|\psi\rangle$. This is the relation between the corresponding observables: 
\begin{eqnarray}
\label{ct1}
& \tilde{X} = X_{1,\theta} =  r X_{\theta} + t X_{0,\theta} , & \nonumber \\
& \tilde{Y} = Y_{2,\theta} = t Y_{\theta} - r Y_{0,\theta} , &
\end{eqnarray}
where $X_{0,\theta}$ and $Y_{0,\theta}$ are the corresponding rotated quadratures for the input mode in vacuum, while $X_{\theta}$ and $Y_{\theta}$ are the rotated quadrature in the signal mode, with:
\begin{eqnarray}
\label{ct2}
 & X_{\theta} = X \cos \theta  + Y \sin \theta   , & \nonumber \\
& Y_{\theta} = - X \sin \theta  + Y \cos \theta  . &
\end{eqnarray}
The quadratures are defined as the real and imaginary parts of the corresponding complex-amplitude operator $a= X+ i Y$ and when necessary we will take advantage of the fact that the vacuum is invariant under quadrature rotations.

\begin{figure}
\centering
\includegraphics[width=6cm]{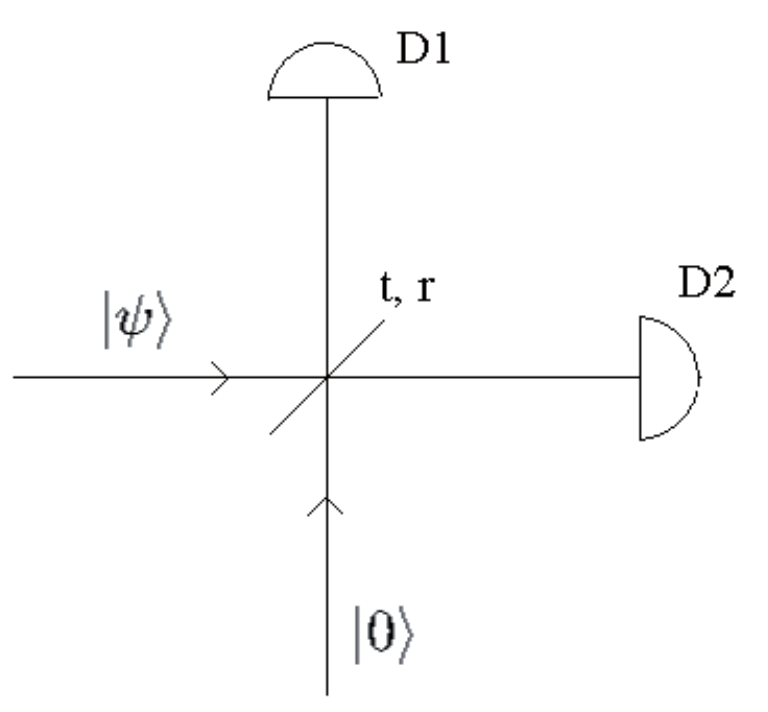}
\label{experimento}
\caption{Diagram of the experimental realization. Detectors D1 and D2 are homodyne detectors measuring quadratures $X_{1,\theta}$ and $Y_{2,\theta}$, respectively.}
\end{figure}

Focusing on the characteristics-based approach in Sec. IIB we begin with the observed joint characteristics for the observables $X_{1,\theta}$ and $Y_{2,\theta}$ 
\begin{equation}
\label{car0}
\tilde{C}^\prime_{X,Y} (u^\prime, v^\prime)  = \langle\,e^{i (u^\prime \tilde{X} +v^\prime \tilde{Y})} \rangle ,
\end{equation}
and we proceed to retrieve the joint characteristics for the observables $X$ and $Y$. Using relations (\ref{ct1}) and (\ref{ct2}) in Eq. (\ref{car0}) 
we consider that the  characteristic function already adapted for our target variables $X$, $Y$ is $\tilde{C}_{X,Y} (u, v) = \tilde{C}^\prime_{X,Y} (u^\prime, v^\prime) $
where
\begin{eqnarray}
\label{csm}
&u = u^\prime \,r\cos\theta - v^\prime t \sin\theta ,& \nonumber \\ 
& v = u^\prime r \sin \theta + v^\prime t \cos\theta . & 
\end{eqnarray}
With this we get that the observed joint characteristic function can be expressed as,
\begin{equation}
\label{car}
\tilde{C}_{X,Y} (u, v) =C^{(S)}_{X,Y} (u, v)\, H_{X,Y} (u, v) ,
\end{equation}
where  
\begin{equation}
C^{(S)}_{X,Y} (u, v) = \langle \psi | e^{i (uX +vY)} | \psi \rangle ,
\end{equation}
\begin{equation}
\label{Hzw}
H_{X,Y}(u,v)=  \langle 0|e^{i(zX_{0,\theta}+wY_{0, \theta})}|0\rangle = e^{-(z^2+w^2)/8} ,
\end{equation}
and
\begin{eqnarray}
\label{zw}
& z = u^\prime t = \frac{t}{r} \left ( u \cos \theta + v \sin \theta \right ), & \nonumber  \\ 
& w=-v^\prime r  = - \frac{r}{t} \left (- u \sin \theta + v \cos \theta \right ) . &
\end{eqnarray}
It  turns out that $C^{(S)}_{X,Y} (u, v)$ is the symmetrically ordered characteristic function for $X$ and $Y$, while $H$ is 
the two-dimensional frequency response with
\begin{equation}
\label{HH}
H_X (u) = H_{X,Y} (u,0), \qquad H_Y (v) = H_{X,Y} (0,v) ,
\end{equation}
and after Eqs. (\ref{Hzw}) and (\ref{zw})
\begin{equation}
H_{X,Y}(u,v) = e^{-(f u^2+gv^2+2 \gamma u v )/8} ,
\end{equation}
with 
\begin{eqnarray}
& f = \frac{t^2}{r^2} \cos^2 \theta + \frac{r^2}{t^2} \sin^2 \theta ,& \nonumber \\
& g = \frac{t^2}{r^2} \sin^2 \theta + \frac{r^2}{t^2} \cos^2 \theta ,& \nonumber \\
& \gamma =  \frac{t^2-r^2}{2 t^2 r^2} \sin (2\theta) . &
\end{eqnarray}

Finally we arrive at the general relation for arbitrary input $| \psi \rangle$ using Eqs. (\ref{fc}), (\ref{car}),  and (\ref{HH}) ,
\begin{equation}
\label{final}
C_{X,Y} ( u,v ) = C^{(S)}_{X,Y} (u, v) \frac{H_{X,Y} (u,v) }{H_{X,Y}(u,0) H_{X,Y} (0,v)} ,
\end{equation}
with
\begin{equation}
\frac{H_{X,Y} (u,v) }{H_{X,Y}(u,0) H_{X,Y} (0,v)}  = \exp \left ( - \gamma u v /4 \right ) .
\end{equation}
This is the factor which makes the whole difference between classical and quantum physics (see discussion below). It holds provided that the input beam splitter is unbalanced $t \neq r$ and that there is a rotation between the measured and inferred variables $\theta \neq 0, \pi/2$. These are the key ingredients allowing the entanglement of statistics required to disclose nonclassical properties, as discussed in Sec. II. 

\subsection{Glauber coherent states} 

To illustrate this procedure with a meaningful case let us assume that $| \psi \rangle$ is a coherent state $| \psi \rangle = | \alpha \rangle$, with 
$a |\alpha \rangle = \alpha | \alpha \rangle$, being $\alpha=x_{0}+i y_{0}$, so that 
\begin{equation}
\label{CScs}
C^{(S)}_{X,Y} (u, v) = \langle\alpha|e^{i(uX+vY)}|\alpha\rangle= e^{i(u x_{0}+v y_{0})} e^{-(u^2+v^2)/8} .
\end{equation}
The final  $C_{X,Y} ( u,v )$ in Eq. (\ref{final}) can be expressed in matrix form as 
\begin{equation}
\label{CM}
C_{X,Y} (u, v) =e^{i\bm{\xi}^{*} \cdot \bm{s}} e^{-\bm{\xi}^{*}M\bm{\xi}} ,
\end{equation}
with $\bm{\xi} = (u,v)^T$, $\bm{s}= (x_0, y_0)^T$, where the subscript $T$ denotes transposition, and $M$ is the $2 \times 2$ real symmetric matrix
\begin{equation}
M = \frac{1}{8} \left( \begin{array}{cc} 1  &  \gamma \\
   \gamma  & 1\end{array} \right) ,
\end{equation}
that does not depend on $\alpha$. The condition for the existence of the integral (\ref{fi}) leading to the $p_{X,Y} (x,y)$ distribution is that  $M$ 
should  be non negative, that is, with positive eigenvalues, which holds provided $| \gamma|  \leq 1$. Otherwise, there is no joint distribution 
$p_{X,Y} (x,y)$, contrary to the classical case  shown in Sec. IIC. Since $M$ dos not depends on $\alpha$ the condition $| \gamma | >1$ leading 
to a nonclassical result can be satisfied at once for every coherent state by a suitable choice of beam splitter and phase $\theta$. For example 
for $\theta= \pi/4$ this is that $t^2 > 1/\sqrt{2}$.

\subsection{Thermal states} 

This result can be extended to mixed thermal states using their expansion in the coherent-state basis as
\begin{equation}
\rho = \frac{1}{\pi \bar{n}} \int d^2 \alpha e^{-|\alpha |^2 / \bar{n}} | \alpha \rangle \langle \alpha |,
\end{equation}
where $\bar{n}$ is the mean number. After the result (\ref{CScs}) we get the following symmetrical-order joint characteristic function 
for thermal states 
\begin{equation}
C^{(S)}_{X,Y} (u, v) = \mathrm{tr} \left [ \rho e^{i(uX+vY)} \right ] = e^{-(1+2 \bar{n})(u^2+v^2)/8} ,
\end{equation}
leading to a final $C_{X,Y} ( u,v )$ of the form (\ref{CM}) with $\bm{s} = \bm{0}$ and 
\begin{equation}
M = \frac{1}{8} \left( \begin{array}{cc} 1+2 \bar{n} &  \gamma \\ \gamma  & 1+2 \bar{n} \end{array} \right) ,
\end{equation}
that depends on the particular thermal state being considered. In this case $M$ fails to be nonnegative when $| \gamma | > 1+2 \bar{n}$. 
This is a more stringent condition as $\bar{n}$ grows, that is as $\rho$ becomes proportional to the identity matrix.  So for every $t,r$ there 
are thermal states with large enough $\bar{n}$ that behave as classical-like. Vice versa, for every $\bar{n}$ we can find $t,r$ values so that 
the thermal state behaves as non classical.

\subsection{Discussion}

\subsubsection{Squeezed $Q$ function}

Although the above analysis focuses on characteristic functions, it may be worth showing that the observed joint statistics
$\tilde{p}_{X,Y} (x,y)$ results from projection of the observed state $| \psi \rangle$ on quadrature squeezed states $| \xi_{x,y} \rangle$ 
\begin{equation}
\tilde{p}_{X,Y} (x,y) = \left | \langle \xi_{x,y} | \psi \rangle \right |^2 ,
\end{equation}
where the states $| \xi_{x,y} \rangle$ are defined by the eigenvalue equation
\begin{equation}
\left [ \left ( r^2 - t^2 \right ) a^\dagger e^{i \theta} + a e^{-i \theta} \right ] | \xi_{x,y} \rangle = 2 \left ( r x+ i t y \right ) | \xi_{x,y} \rangle .
\end{equation}
This can be easily shown from the defining eigenvalue equations 
\begin{equation}
\tilde{X} |\phi \rangle = x  |\phi \rangle, \quad \tilde{Y} |\phi \rangle = y  |\phi \rangle ,
\end{equation}
combining them as 
\begin{equation}
\left ( r \tilde{X} + i t \tilde{Y}  \right )|\phi \rangle = \left ( r x+i t y \right )  |\phi \rangle ,
\end{equation}
using Eqs. (\ref{ct1}) and (\ref{ct2}), and then finally projecting on the vacuum on the mode $a_0$, being $| \xi_{x,y} \rangle = \langle 0 | \phi \rangle$. 

Thus the measuring states $| \xi_{x,y} \rangle$ are quadrature squeezed states provided that the input beam splitter is unbalanced $t \neq r$.
The squeezing direction in the $X$, $Y$ plane is specified by the phase $\theta$. This is to say that the statistics $\tilde{p}_{X,Y} (x,y)$ is actually 
a squeezed $Q$ function. This reduces to the standard $Q$ in the balanced scheme $t=r$ so that $| \xi_{x,y} \rangle$ become coherent states 
$| \alpha \rangle$ with $\alpha = (x+iy)/\sqrt{2}$ \cite{Q}.

\subsubsection{Nonclassical measurement}

The fact that the statistics is given by projection on nonclassical states does not spoil the interest of the result. Actually, this is the same 
case of the most paradigmatic nonclassical tests, such as subPoissonian statistics and quadrature squeezing. They also crucially rely 
on the projection on highly non classical  measuring  states: number states and infinitely squeezed states, respectively. Moreover, it has been shown 
that such nonclassical effects vanish if the measuring states become classical-like \cite{LU17}. 

\subsubsection{The vacuum}

For the proper comparison with the classical model in Eq. (\ref{jj}) it must be understood that in this case we refer to classical models where 
the vacuum means a field of definite zero amplitude. Since our result relies on the frequency response (\ref{Hzw}) it may be regarded as 
a quantum-vacuum effect, as other relevant nonclassical effects in quantum optics such as spontaneous emission \cite{P}.  

\subsubsection{Entanglement of statistics}

We think it is worth pointing out that the nonclassical test found here, that is $M$ lacking positive semidefiniteness,  has a very close 
resemblance with the inseparability criterion for Gaussian states \cite{ec}. This might be expected since we have already commented on the fact that 
nonclasicality is equivalent to the lack of factorization for the observed statistics.  

\bigskip

\section{Conclusion}

We have used a simple and general protocol to disclose nonclassical effects for states customarily 
regarded as   the most classical states.   These are the Glauber coherent states, thermal 
states, and the  SU(2) coherent states  for spin variables.  Moreover, we have shown that for all states  
there is always a measurement setting where the inferred joint distribution cannot represent probabilities, with 
the only exception being the totally incoherent mixed state in finite-dimensional spaces.  So there is no state that 
would always allow us to infer true probability distributions.   These results are consistent with previous 
works that have also reported nonclassical properties for these  states following different approaches 
\cite{thisone,ncpn},  and with some more recent works extending nonclassical correlations and entanglement 
to all quantum states \cite{qq}. 

We have shown that nonclassicality holds because the observed joint probability distribution is not separable. 
We have to stress that this does not refer to actual particles, but just to the dependence of the statistics 
on the two observed variables. 
 
\bigskip
 
\section*{ACKNOWLEDGMENTS}

A. L.  thanks Profs. A. R. Usha Devi, H. S. Karthik, A. Iosif, and M. Acedo for valuable comments, as well as encouraging 
support from Ceres team. L. M. gratefully thanks a Collaboration Grant from the Spanish Ministerio de Educaci\'{o}n, 
Cultura y Deporte. A. L. acknowledges financial support from Spanish Ministerio de Econom\'ia y Competitividad 
Projects No. FIS2012-35583 and FIS2016-75199-P, and from the Comunidad Aut\'onoma de Madrid research  consortium 
QUITEMAD+ Grant No. S2013/ICE-2801.

\end{document}